\documentclass[12pt]{JHEP3}

\newcommand{\tfrac}[2]{{\textstyle \frac{#1}{#2}}}
\newcommand{\ZZ}{\mathbb{Z}}
\newcommand{\nsp}{\!\!\!\!\!\!}

\title{Fermionic Zero Modes of Supergravity Cosmic Strings}

\author{Philippe Brax \\
Service de Physique Th\'eorique, CEA/DSM/SPhT,
Unit\'e de recherche associ\'ee au CNRS,
CEA--Saclay, F--91191 Gif/Yvette cedex, France \\
E-mail: \email{brax@spht.saclay.cea.fr}}
\author{Carsten van de Bruck \\
Department of Applied Mathematics, University of Sheffield,
Hounsfield Road, Sheffield, S3 7RH, UK \\
E-mail: \email{C.vandeBruck@sheffield.ac.uk}}
\author{Anne-Christine Davis \\
DAMTP,Centre for Mathematical Sciences, University of 
Cambridge, \\ Wilberforce Road, Cambridge, CB3 0WA, UK \\
E-mail: \email{A.C.Davis@damtp.cam.ac.uk}}
\author{Stephen C. Davis \\
Instituut-Lorentz for Theoretical Physics, Postbus 9506,
NL--2300 RA Leiden, The Netherlands \\
E-mail: \email{sdavis@lorentz.leidenuniv.nl}}

\abstract{
Recent developments in string theory suggest that cosmic strings
could be formed at the end of brane inflation. Supergravity provides
a realistic model to study the properties of strings arising in
brane inflation. Whilst the properties of cosmic strings in flat
space-time have been extensively studied there are significant
complications in the presence of gravity. We study the effects of
gravitation on cosmic strings arising in supergravity. Fermion zero
modes are a common feature of cosmic strings, and generically occur
in supersymmetric models. The corresponding massless currents can
give rise to stable string loops (vortons). The vorton density in
our universe is strongly constrained, allowing many theories with
cosmic strings to be ruled out. We investigate the existence of
fermion zero modes on cosmic strings in supergravity theories. A
general index theorem for the number of zero modes is derived. We
show that by including the gravitino, some (but not all) zero modes
disappear. This weakens the constraints on cosmic string models.
In particular, winding number one cosmic D-strings in models of
brane inflation are not subject to vorton constraints. We also
discuss the effects of supersymmetry breaking on cosmic D-strings.
}


\preprint{hep-th/0604198}

\begin{document}

\section{Introduction}

There has been a resurgence of interest in cosmic strings arising
from recent results in fundamental superstring theory (for reviews 
see~\cite{polchinski,kibble}). Indeed, in string theory a generic feature
of D-brane anti-brane annihilation is the production of lower dimensional
branes, with $D3$ and $D1$ branes, or D-strings, favoured~\cite{mahbub}.
Consequently, in models of brane inflation where a period of inflation is
caused by the attraction, and subsequent annihilation, of a D-brane and
anti-brane, D-strings form naturally~\cite{tye}. Depending on the details of
the theory, fundamental strings, or F-strings can also arise~\cite{copeland}
and, in certain classes of models, axionic local strings~\cite{axion}.
The D-strings arising in such models have many features in common with
cosmic strings formed in a cosmological phase transition (for reviews 
see~\cite{vilenkin&shellard,hindmarsh&kibble}). However, there are
added complications arising from supergravity which need to be addressed.

Cosmic strings in supersymmetric theories have been investigated and
shown to give rise to two sorts of strings, called D-term and F-term
strings~\cite{susyCS}, where the D and F refer to the type of
potential required to break the symmetry. For the D-term case, this
has been extended to supergravity~\cite{edelstein}, and an
analysis of the BPS configurations led to the conjecture that the
D-term strings of supergravity were in fact
D-strings~\cite{dvali}. Consequently, an analysis of strings in
supergravity theories should provide insight into those formed at the
end of brane inflation.

Strings arising in supersymmetric theories have fermion zero modes in
the string core~\cite{susyCS}. This is an inevitable result of the
couplings and particle content required by the supersymmetry algebra.
In certain cases the zero modes survive supersymmetry
breaking~\cite{susybreak}. The presence of fermion zero modes changes
the cosmology of cosmic strings drastically since they result in the
string carrying a current\footnote{Such strings are widely referred to as
`superconducting', when in fact they are actually just perfect
conductors. To be superconducting they would not only need fermion
zero modes, but the fermions would also have to form into Cooper pair
bound states~\cite{cooper}. For cosmology the distinction between
super- and perfect conductivity is not very important, since most of
the cosmological properties of conducting strings follow from the fact
that the currents are conserved (which is true for both types of current).}.
Ordinary cosmic strings are either long
strings or closed loops. The latter usually decay via the emission of
gravitational radiation, resulting in a scaling solution of the
string network. However, when the string carries a current the loop
can be stabilised by the current carriers~\cite{vortons}. These
stable loops, or vortons, put constraints on the underlying
theory~\cite{BCDT,CD}. The presence of zero modes on cosmic strings
has been well studied in the case of global
supersymmetry~\cite{susyCS}. An important question to ask is whether
these results go over to the case of supergravity. An analysis of
supersymmetry transformations in ref.~\cite{rachel} suggests that
conductivity is reduced for supergravity cosmic strings.
In this paper we will address the question in detail.

In general supergravity contains both D- and F- terms, and cosmic
strings can arise from both classes of theory, though only the
former have been studied to date. Here we investigate both D- and
F-term theories. This is particularly relevant since general brane
inflation models contain both types of terms. Of course, only
strings arising from the D-term would be BPS states. The paper is
arranged as follows. In section~\ref{sec:cs} we consider the
gravitational background created by cosmic strings. Here we solve
for the most general metric appropriate to cylindrical symmetry, which
will be used later when calculating string zero modes. In
section~\ref{sec:ind} we prove a general index theorem for the Dirac
operator in the cosmic string background including gravitation. We
consider the most general form of mass matrices that can arise in
such theories. Supersymmetric examples are reviewed in
section~\ref{sec:susyex}, where we apply the index theorem to D- and
F-term strings. In section~\ref{sec:sugra}, we consider the
supergravity case. Here we map the gravitino equations onto Dirac
equations amenable to a treatment as in section~\ref{sec:ind}.
Supergravity examples are considered in section~\ref{sec:sugex}. We
find that the presence of the gravitino generically reduces the number
of zero modes on supergravity cosmic strings. 
In particular, we find that for BPS D-strings the
number of chiral zero modes are reduced from $2n$ in the global case
to $2(n-1)$ in supergravity. Thus winding number $1$ D-strings evade
the stringent vorton constraints found for chiral
theories~\cite{CD}. In section~\ref{sec:curr}, we extend our results on zero
modes to the issue of massless currents on the cosmic string
world--sheet. We conclude in section~\ref{sec:conc}.

\section{Gravitating cosmic strings}
\label{sec:cs}

We consider a cosmic string configuration created by the spontaneous
breaking of gauge symmetries. The simplest example of this would be a
$U(1) \to I$ breaking, although any symmetry breaking whose vacuum
manifold is not simply connected can produce cosmic string
solutions. In the cosmic string background, the scalar fields have a profile
\begin{equation}
\phi^i(r,\theta)= e^{i\theta T_s} \phi^i(r)
\end{equation}
where the string generator $T_s$ is some linear combination of
generators, $T^u$, from the broken gauge group. In this paper we
will define it to include the winding number of the string. The
covariant derivative of the scalar fields is
\begin{equation}
{D}_\mu \phi^i =(\partial_\mu - i T^u A^u_\mu )\phi^i \, .
\end{equation}
This must vanish at infinity to ensure a finite energy solution.
This is achieved by having
\begin{equation}
T^u A^u_\theta\vert_{r \to \infty} = T_s \, .
\end{equation}
The simplest string gauge fields will only involve one generator,
although more generally solutions can have several gauge fields,
each with it own length scale. When there is only one generator, $T_s
= nQ$ where $Q$ is the electric charge and $n$ is the string's winding number.

The gravitational effects of a cosmic string lead to a deficit angle
in the far away metric of spacetime. In the following we
will consider the metric
\begin{equation}
ds^2=e^{2 B}(-dt^2 + dz^2 + d\rho^2 + \alpha^2 d\theta^2)
\end{equation}
for a cosmic string configuration. This is the most general
cylindrically symmetric metric, as discussed by
Thorne~\cite{Thorne}. Notice that $B$ is an overall conformal factor
and $\alpha$ is related to the deficit angle of the cosmic strings.
The Einstein equation involves the energy momentum tensor and reduces to
\begin{eqnarray}
\alpha''&=& \kappa \alpha e^{2B} (T^t_t+ T^\rho_\rho)\nonumber \\
2B''&=& -(B')^2+\kappa T^\theta_\theta\nonumber\\
(\alpha B')'&=& \frac{\kappa}{2} \alpha e^{2B}
(T^\rho_\rho+T^\theta_\theta) \, .
\end{eqnarray}
These equations can be solved numerically by imposing that the metric
is regular at the origin and satisfies $B'(0)=0$, $\alpha(0)=0$ and
$\alpha'(0)=1$. We also impose $B(\infty)=0$.

In order to determine of the number of fermion zero modes we will only
require the form of the string solution at large and small $r$.
Hence an approximate solution will be sufficient for our analysis.
We will look for solutions using the top hat approximation whereby
the scalars are assumed to be constant inside and outside the cosmic
string with a jump at the string radius. Let us consider the
solution inside. First of all, notice that the energy momentum
tensor is almost constant with $T^\rho_\rho\approx T^\theta_\theta$.
Let us define $m^2= \kappa (T^t_t+T^\rho_\rho)$, then the metric
inside reads
\begin{equation}
\alpha= \frac{\sin(m\rho)}{m}
\end{equation}
while
\begin{equation}
B=- \frac{\kappa T^\rho_\rho}{m^2} \left(\ln \sin m\rho - m \int^{\rho}
\frac{d\rho'}{\sin m \rho'}\right) + B_* \, .
\end{equation}
A good approximation for small $\rho$ is given by
\begin{equation}
 B\approx
\frac{\kappa}{4} T^\rho_\rho \rho^2 + B_*
\end{equation}
satisfying the boundary conditions at the origin. Let us now
consider the outside solution. In this region, the energy momentum
tensor is approximately zero and therefore $B$ is zero. Now we also
find that
\begin{equation}
\alpha=C_0 + C_1 \rho \, .
\end{equation}
When $C_1\ne 0$, the solution is a cosmic string solution with a
deficit angle $\Delta= 2\pi (1-C_1)$. Indeed, after defining $\tilde
\rho= (\rho+C_0/C_1)$ the metric becomes
\begin{equation}
ds^2=-dt^2 + dz^2 + d\tilde \rho^2 + C_1 ^2 \tilde \rho^2 d\theta^2
\end{equation}
near infinity.

Using the change of variables $dr = e^B d\rho$ we obtain
\begin{equation}
ds^2=e^{2 B}(-dt^2 +dz^2) + dr^2 + C^2 d\theta^2
\end{equation}
which is the metric we will use for the rest of this paper. Near
the origin $C \approx r$ and far from the string $C = C_1 r + C_0
+ O(1/r)$.

In supersymmetric theories, a cosmic string breaks all supersymmetries
in its core in general. BPS objects are an exception to this rule,
as they leave 1/2 of the original supersymmetry unbroken.
D-strings, in which we will be interested in this paper, are an example
of this. These strings have vanishing $T^\rho_\rho$, and the
conformal factor $B$ is identically zero. As we will discuss in more
detail in subsection~\ref{ssec:Lag}, the string will only be BPS if
the superpotential vanishes for the string solution. The only
non-zero potential energy for the string comes from the D-term
corresponding to $T_s$.

Let us characterise the BPS cosmic strings. We are considering a
$U(1)$ symmetry breaking, so we take $T_s \phi^i = n Q_i \phi^i$ and
$A_\mu = \delta^\theta_\mu n a(r)$. The bosonic fields satisfy first
order equations
\begin{equation}
\partial_r \phi^i = \mp n\frac{1-a}{C} Q_i \phi^i
\label{Dstr1}
\end{equation}
and
\begin{equation}
\mp n \frac{\partial_r a}{C} = D = \xi - \sum_i Q_i K_i \phi^i
\end{equation}
where $\xi$ is the Fayet-Iliopoulos term which triggers the breaking of
the $U(1)$ gauge symmetry and $K$ is the Kahler potential. The
Einstein equations reduce to $B'=0$ and
\begin{equation}
C'=1\pm A^B_\theta
\label{Dstr3}
\end{equation}
where
\begin{equation}
A^B_\mu = \frac{i}{2} (\bar K_{\bar \jmath} D_\mu \bar \phi^{\bar
\jmath} - K_j D_\mu \phi^j) +\xi A_\mu \, .
\label{AB}
\end{equation}
The simplest BPS configuration will just have one $\phi$, with unit
charge. Notice that BPS cosmic strings are solutions of first order
differential equations. These equations are consequences of the
Killing spinor equations when requiring the existence of 1/2
supersymmetry.

In order to make sure that the deficit angle of
the string is positive ($C_1<1$), we must choose $\pm = -\mathrm{sign} (n)$,
which gives
\begin{equation}
C_1= 1 - |n|\xi \, .
\end{equation}
Notice that the deficit angle of the string is directly proportional
to the winding number. Restricting ourselves to positive tension
strings implies that the winding $n$ cannot be arbitrarily large.

We will also consider the non-BPS configurations where the tension is
higher than the BPS tension, this implies that $C_1$ is less than the BPS case
\begin{equation}
C_1\le 1- \vert n \vert \xi \, .
\end{equation}
We will see that this change in the deficit angle can alter the
number of fermion zero modes on a string.

\section{Zero modes and the index theorem}
\label{sec:ind}

\subsection{The Dirac equation}

When fermions live in a cosmic string background, they are subject
to both the electromagnetic interaction and the gravitational
interaction. We consider a family of $n_f$ Weyl fermions with a
Lagrangian
\begin{equation}
\mathcal{L}= g_{ij} \bar \chi^{i}_{\dot\alpha} i \bar
\sigma_\mu^{\dot \alpha \alpha} {D}^\mu \chi^{j}_{\alpha}
+\frac{1}{2} \bar \chi^i_{\dot\alpha} M_{ij} \bar \chi^{j\dot
\alpha} + (\mathrm{c.c.})
\end{equation}
where $g_{ij}(\phi,\bar\phi)$ is a sigma-model metric depending on the cosmic
string background. The covariant derivative involves the gauge
fields $A^a_\mu$ and the spin connection term $w_\mu$
\begin{equation}
{D}_\mu\chi^i = \left(\partial_\mu + w_\mu\right)\chi^i
-i A^u_\mu (T^u){}^i_j \chi^j +\Gamma_{jk}^i {D}_\mu \phi^j
\chi^k + J_{\mu j}^i \chi^j
\label{Dchi}
\end{equation}
$w_\mu = -w_{\mu ab}\sigma^{ab}/2$, where $a, b$ are flat indices and
$\sigma^{ab}= (\sigma^a \bar \sigma^b -\sigma^a \bar \sigma^b)/4$.
In particular we find that $\sigma^{12}=-i\sigma^3/2$.

For a cosmic string, $w_{\mu ab}$ has only three non-zero
entries
\begin{equation}
w_{t\hat r \hat t} = w_{z \hat r \hat z} = -B'
\, , \qquad w_{\theta \hat \theta \hat r}= C'
\end{equation}
where hatted quantities are flat indices and we identify $\hat t
\equiv 0$, $\hat r \equiv 1$, $\hat \theta \equiv 2$ and $\hat
z\equiv 3$.

We have also included the Levi-Civita connection $\Gamma^i_{jk}$
to take into account the reparametrisation of the sigma-model
scalar fields. Any other contributions are included in $J_\mu$.
Given the form of the string solution, we will assume $J_z$ and
$J_t$ are zero. For the supergravity D-strings discussed in
section~\ref{sec:cs} we find that
\begin{equation}
J_\mu=\frac{1}{4}(K_j D_\mu \phi^j-\bar K_{\bar \jmath} D_\mu \bar
\phi^{\bar \jmath})
\end{equation}
and only $J_\theta$ is non-vanishing.

Notice that the Weyl fermions live in the spin-bundle of the
sigma-model manifold defined by the scalar field, hence we have as
many fermions as scalars in the curved case. We can have more
fermions than scalar fields by considering fermions having a flat
sigma-model metric. This happens in the supersymmetric setting with
the Kahler manifold defining the set of scalar fields. The gauginos
are not associated to the scalar fields and their sigma-model metric
is flat.

In the following we will diagonalise the sigma-model
metric using the vielbein $e^a_i$ such that
\begin{equation}
g_{ij}= \delta_{ab} e^a_i e^b_j
\end{equation}
and redefine the fermions
\begin{equation}
\chi^i=e^i_a \chi^a \, .
\end{equation}
We also choose the $\chi^a$ to diagonalise the string generator, so that
$T_s \chi^a = q_a \chi^a$.

The Lagrangian can then be written as
\begin{equation}
{\cal L}= \bar \chi^a i \bar \sigma^\mu D_\mu \chi^a +\frac{1}{2}
\bar \chi^a M_{ab} \bar \chi^b + (\mathrm{c.c.})
\end{equation}
where summation over the $a,b$ indices is understood and
\begin{equation}
M_{ij}= M_{ab} e^a_i e^b_j \, .
\end{equation}
The covariant derivative becomes
\begin{equation}
D_\mu\chi^a \to D_\mu\chi^a
+ e^a_j \left(\frac{\partial}{\partial \phi^k} \right)
e^j_c D_\mu \phi^k \chi^c
+ e^a_j \left(\frac{\partial}{\partial \bar \phi^{\bar k}} \right)
 e^j_c D_\mu \bar \phi^{\bar k} \chi^c
\, .
\end{equation}
For a wide range of models this cancels the term arising from
the Levi-Civita connection $\Gamma^i_{jk}$ of the scalar manifold,
although this is not always the case.
An example where such a cancellation does occur has been
given in ref.~\cite{axion} for a Kahler metric
$g_{ij}=\partial_i\partial_{\bar j} K$, where $K=-\ln (S+\bar S)$ is
the Kahler potential of the dilaton. In this case the two
contributions from the non-flatness of the sigma model metric
cancel in the cosmic string background. More generally, if any part of the
connection is not cancelled, we absorb it into $J_\mu$.

The zero modes can be separated into $\sigma^3$ eigenstates. The upper
component spinors will be denoted by $\chi^a_+$ (for positive chirality),
and the lower ones by $\chi^a_-$ (for negative chirality). The Dirac
equations now read
\begin{equation}
\left(\partial_r + J_r + \frac{i}{C}\left[\partial_\theta
-\frac{i}{2}C'-i T^u A^u_\theta + J_\theta \right]\right)\chi^a_+
-iM_{ab}(\chi^b_+)^*=0
\label{dirind1}
\end{equation}
and similarly
\begin{equation}
\left(\partial_r + J_r -\frac{i}{C}\left[\partial_\theta
+\frac{i}{2}C'-iT^u A^u_\theta + J_\theta\right]\right)\chi^a_-
+iM_{ab}(\chi^b_-)^*=0 \, .
\label{dirind2}
\end{equation}
Notice that the spin connection parts along the $z$ and $t$
directions cancel. We will look for normalisable solutions of
these equations, and use the natural choice of norm
\begin{equation}
\vert\vert \chi \vert\vert^2
= \sum_a\int dr d\theta C e^{2B} \vert \chi^a \vert^2 \, .
\end{equation}
As the $B$ factor is constant at the origin and at infinity, we
see that finiteness of the norm is equivalent to $L^2$ normalisability
of the $\chi_\pm^a$.

Gauge invariance implies that the mass matrix must be factorisable
\begin{equation}
M_{ab}=e^{i(q_a +q_b)\theta}\mathcal{M}_{ab}(r) \, .
\end{equation}
We can assume that both $M_{ab}$ and $J^a_{\mu b}$ are not
simultaneously block diagonal. If they are, we simply separate the
fermions into several independent systems, and analyse each one
individually.

As the mass matrix is obtained using algebraic combinations of the
bosonic fields, it is a single-valued matrix. In general this
implies that the fermion charges $q_a$ must all be either integers
or half integers, although there are some exceptions which require a
different treatment. One exception is if the mass matrix is block
off-diagonal. By this we mean that the only non-zero entries are in
the $n_1\times n_2$ upper-right and the $n_2\times n_1$ lower left
corners of the matrix (with $n_1+n_2=n_f$). This will be the case
if system has only Dirac mass terms. It is also relevant for
D-strings. Another exception to the above rule is if $M_{ab} \equiv
0$. In both these cases the charges are not automatically integers or
half integers. We will deal with them in later subsections.

\subsection{Generic mass matrix}
\label{ssec:in1}

The index theorem giving the number of normalisable zero modes is
obtained by analysing the normalisability of the modes at the
origin and at infinity. Close to the origin the metric is simply
the flat metric in polar coordinates. One can decompose the
fermions using the ansatz
\begin{equation}
\chi^a_\pm= \frac{1}{\sqrt{C}} e^{i q_a\theta}
\left\{U^a_\pm(r) e^{\pm il\theta} \mp
V^{a*}_\pm(r) e^{\mp il\theta}\right\} \, . \label{UVdef}
\end{equation}
We require that the fermions are single valued when $\theta \to
\theta +2\pi$ around the origin. This implies that
\begin{equation}
q_a + l, q_a - l \in \ZZ+\frac{1}{2} \, .
\end{equation}
These two conditions are satisfied provided $l$ is an integer when
the $q_a$ are half-integer, or the converse. We will now use the
approximate large and small $r$ behaviour of the fermion field
equations to determine the number of zero modes.

Close to the origin $A_\theta^u T^u$ vanishes, so to leading order
\begin{equation}
\partial_r U^a_\pm
- \frac{1}{r}\left[l \pm q_a \right]U^a_\pm +i \mathcal{M}_{ab}
V^b_\pm =0 \label{U0}
\end{equation}
and
\begin{equation}
\partial_r V^a_\pm
+ \frac{1}{r}\left[l \mp q_a \right] V^a_\pm + i\mathcal{M}_{ab}
U^b_\pm =0 \, . \label{V0}
\end{equation}
We have also assumed that $J_\mu$ is subdominant near the origin, as
it is frequently the case in supergravity. If it is not subdominant,
the results we are deriving will not be valid (although the approach
we are using can be generalised).

Now since $\mathcal{M}_{ab} =O(1)$, this implies that the solutions behave as
\begin{equation}
U^a_\pm \sim r^{ \pm q_a +l} \, , \qquad V^a_\pm \sim r^{\pm q_a-l}
\label{UV0}
\end{equation}
where we have again assumed that $J_\mu$ can be neglected.

Normalisability of the above $\chi^a_\pm$ solutions at small $r$
demands that either $l\ge \mp q_a+ 1/2$ or $l\le \pm q_a -1/2$ for
a given $l$. The total number of normalisable, small $r$
solutions for a given $l$ is then
\begin{equation}
N_0^\pm(l)=\sum_{i=1}^{n_f}\left\{ I\left(l\le \pm q_a-\tfrac{1}{2}\right)
+ I\left(l\ge \mp q_a+\tfrac{1}{2}\right)\right\}
\label{N01}
\end{equation}
where $I$ is the characteristic function of each interval [so
$I(\mathrm{true})=1$ and $I(\mathrm{false})=0$].

At infinity the behaviour of the fermion fields is mainly
determined by the mass matrix. Defining $\tilde U, \tilde V$ to be
linear combinations of $U$ and $V$ which diagonalise
$\mathcal{M}_{ab}$ at $r=\infty$, the asymptotic Dirac equation
reads
\begin{equation}
\partial_r \tilde U^a_\pm
- \frac{l}{C_1r}\tilde U^a_\pm +i \lambda_a \tilde V^a_\pm=0
\label{Ui}
\end{equation}
\begin{equation}
\partial_r \tilde V^a_\pm
+ \frac{l}{C_1r} \tilde V^a_\pm + i\lambda_a \tilde U^a_\pm=0
\label{Vi}
\end{equation}
where $\lambda_a$ are the eigenvalues of the mass matrix. If all
$\lambda_a$ are non-zero, the $2n_f$ solutions will have
exponential behaviour, and exactly half of them vanish at infinity,
and so be normalisable.

On the other hand if some $\lambda_a=0$ then the leading order behaviour
of those solutions is instead
\begin{equation}
\tilde U^a_\pm \sim r^{l/C_1} \, , \qquad
\tilde V^a_\pm \sim r^{-l/C_1} \, .
\label{UVi}
\end{equation}
One of these will be normalisable if $|l| > C_1/2$. The total number
of normalisable, large $r$ solutions for a given $l$ is therefore
\begin{equation}
N^\pm_\infty(l) = n_f - n_z I\left( |l| \leq \tfrac{C_1}{2}\right)
\label{Ni1}
\end{equation}
where $n_z$ is the number of massless fermions at infinity, or
equivalently the number zero eigenvalues $\lambda_a$. It will be
useful to express eq.~(\ref{Ni1}) in a similar form to
eq.~(\ref{N01}). If we define
\begin{equation}
\tilde q_\pm = \left\{ \begin{array}{l@{\ \ }l}
 \mp \frac{1}{2}& \mathrm{if} \ q_a \in \ZZ +1/2 \\
0 & \mathrm{if} \ q_a \in \ZZ \\
\end{array} \right.
\label{qtdef}
\end{equation}
we can write
\begin{equation}
N_\infty^\pm(l)=n_f -n_z + n_z \left\{
I\left(l\le \pm \tilde q -\tfrac{1}{2}\right)
+ I\left(l\ge \mp \tilde q +\tfrac{1}{2}\right)\right\} \, .
\end{equation}
We have explicitly taken into account the fact that $0<C_1<1$ for a
cosmic string with a non-vanishing deficit angle.

The set of normalisable zero modes is characterised by conditions at
the origin and infinity. When the mass matrix is generic, the total
number of zero mode solutions for a given $l$ forms a vector space of
dimension $2n_f$. Solutions which are normalisable at the origin form
an $N_0^\pm(l)$-dimensional subspace of this. The solutions which are
normalisable at infinity form an $N_\infty^\pm(l)$-dimensional
subspace. Hence the number of independent solutions which are
normalisable everywhere is equal to the dimension of the intersection
of these two subspaces. This implies there are
\begin{equation}
N^\pm(l)= [N^\pm_0(l) + N^\pm_\infty(l)- 2n_f]_+
\label{Nl}
\end{equation}
solutions, where $[x]_+=x$ if $x\ge 0$ and zero otherwise. Of course
the above argument only gives a lower bound for the number of zero
modes. It is possible there will be additional solutions, although
this will generally require fine tuning of the string background.
Extra solutions would also occur if the mass matrix $M_{ab}$ has
some degeneracy. For example if it is block diagonal. However we
have already included this possibility in the analysis.

To simplify the above expression~(\ref{Nl}) it is convenient to
separate the winding numbers $q_a$ into $n_+$ positive values
$q^+_a$ and $n_-$ negative values $q^-_a$. In what follows we will
concentrate on the $\chi_a^+$ modes, although a similar analysis
applies for $\chi_a^-$. To take into account the effect of any massless
fermions that are present, we define $\hat q^-_a$ to be the union of
$q^-_a$ and $n_z$ copies of $\tilde q_+$. Note that
$\tilde q_\pm$ has been defined so that, like the $q_a$, it will
either be an integer or a half integer. Then
\begin{eqnarray}
&& N_0^+(l) + N^+_\infty(l) - 2n_f =
\nonumber \\ && \hspace{0.5in}
\sum_{a=1}^{n_+}
I\left(-q^{+}_a+\tfrac{1}{2}\le l \le q^{+}_a-\tfrac{1}{2}\right)
-\sum_{a=1}^{n_-+n_z}
I\left(\hat q^{-}_a-\tfrac{1}{2}<l<-\hat q^{-}_a+\tfrac{1}{2}\right)
\, . \hspace*{0.5in}
\label{Nl1}
\end{eqnarray}

Any zero $q_a$ do not contribute to the sum and therefore do not
lead to zero modes. We will place the charges in decreasing order of
magnitude, so that $q^+_1\ge \dots q^+_{n_+}$ and
$\hat q^-_1 \le \dots \hat q^-_{n_-+n_z}$. This allows the above
expression~(\ref{Nl1}) to be rewritten as
\begin{eqnarray}
\sum_{a=1}^{n_*}
\biggl\{
I\left(-q^{+}_a+\tfrac{1}{2}\le l \le \hat q^{-}_a-\tfrac{1}{2}\right)
+ I\left(-\hat q^{-}_a+\tfrac{1}{2}\le l \le q^{+}_a-\tfrac{1}{2}\right)
\hspace{1in} \nonumber \\
- I\left(\hat q^{-}_a-\tfrac{1}{2}<l<-q^{+}_a+\tfrac{1}{2}\right)
- I\left(q^{+}_a-\tfrac{1}{2}<l<-\hat q^{-}_a+\tfrac{1}{2}\right) \biggr\}
\hspace{0.5in} \nonumber \\
+\! \sum_{a=n_* +1}^{n_+} \! \!
I\left(-q^{+}_a+\tfrac{1}{2}\le l \le q^{+}_a-\tfrac{1}{2}\right)
-\! \sum_{a=n_* +1}^{n_-+n_z} \! \!
I\left(\hat q^{-}_a-\tfrac{1}{2}<l<-\hat q^{-}_a+\tfrac{1}{2}\right)
\label{Nl2}
\end{eqnarray}
with $n_* = \min(n_+,n_-+n_z)$.

Taking into account the double counting due to the $l\to -l$
symmetry and the fact that only one of the solutions is
non-vanishing for $l=0$, we find that the number of real solutions
is given by
\begin{equation}
N^\pm= \sum_l N^\pm(l)
\end{equation}
with $N^\pm(l)$ given by the above expression~(\ref{Nl}). It can be
seen that for a given $l$, the only non-zero terms in the
sums~(\ref{Nl2}) are either all positive or all negative. If they
are all negative, the expression for $N^+(l)$~(\ref{Nl}) evaluates
to zero. Hence we can ignore all negative terms in the
sum~(\ref{Nl2}) when we evaluate $N^+(l)$. Our expression
for the total number of zero modes reduces to a sum of only positive
terms. We can therefore change the
order of the $l$ and $a$ summations. From this we obtain
\begin{equation}
N^+ = 2 \sum_{a=1}^{n_*} \left[q^+_a + \hat q^-_a\right]_+
+ 2 \sum_{a=n_*+1}^{n_+} q^+_a
\label{Np1}
\end{equation}
where $q^{+}_a$ are all the positive $q_a$, and $\hat q^{-}_a$ are
all the negative $q_a$ and $n_z$ copies of $\tilde q_+$. For the
summations we have taken $\sum_{a=a_1}^{a_2}\ldots = 0$ if
$a_1 > a_2$. The above expression is the generalisation of the index
theorem to self-gravitating cosmic strings. It coincides with
previous index theorems~\cite{oldindex,index} which apply when there are no
massless fermions.

Using the fact that $|\tilde q_+| \leq |q^-_a|$, we can rewrite the
above sum as
\begin{equation}
N^+ = 2 \nsp \sum_{a=1}^{\min(n_-,n_+)} \nsp
\left[q^{+}_a+q^{-}_a\right]_+
+ 2 \nsp \sum_{a=n_-+1}^{\min(n_+,n_-+n_z)} \nsp
\left[q^{+}_a+\tilde q_+\right]_+
+ 2 \sum_{a=n_-+n_z+1}^{n_+} \nsp q^{+}_a \, .
\end{equation}

A similar analysis applies to the other chirality. This time the
roles of the $q^{+}_a$ and $q^{-}_a$ are interchanged. If there are
massless fermions present, we define $\hat q^{+}_a$ to be $q^{+}_a$
with $n_z$ copies of $\tilde q_-$ added. Taking
$n'_* = \min(n_-,n_++n_z)$, we obtain
\begin{equation}
N^- = 2 \sum_{a=1}^{n'_*} \left[-\hat q^+_a - q^-_a\right]_+ + 2
\sum_{a=n'_*+1}^{n_-} (-q^-_a) \, .
\end{equation}
From the above expressions, we find that if $n_+=n_-$, the presence
of massless fermions in the vacuum does not affect $N^\pm$.

\subsection{Block off-diagonal or Dirac-like mass matrices}

Now let us extend our results to Dirac-style mass matrices. The only
non-zero elements of $M_{ab}$ are in the $n_1 \times n_2$
upper-right corner and the $n_2 \times n_1$ lower-left corner of the
matrix. The total number of fermions is $n_f= n_1+n_2$.
 The winding operator can be decomposed as
\begin{equation}
T_s= T_s^0 + T_s^\eta
\end{equation}
in such a way that $T_s^0$ has half-integer eigenvalues. Defining
$q^{(1)}_a = q_a$ for $a=1\ldots n_1$ and $q^{(2)}_a = q_{a+n_1}$
for $a=1\ldots n_2$, we can choose some value $\eta$, for which the charges
satisfy $q^{(1)}_a +\eta \in \ZZ+1/2$ and $q^{(2)}_a - \eta \in
\ZZ+1/2$, i.e.\ $T^\eta_s$ has eigenvalues $-\eta$ and
$\eta$ respectively. There is no need for the $q_a$'s to be
integers or half integers. We use the decomposition
\begin{equation}
\chi^a_\pm = \frac{1}{\sqrt{C}} \times \left\{ \begin{array}{ll}
e^{i(q_a \pm l)\theta} U^a_\pm(r) & a = 1 \ldots n_1 \\
e^{i(q_a \mp l)\theta} V^{a*}_\pm(r) & a = (n_f-n_2+1) \ldots n_f
\end{array} \right.
\label{UVd}
\end{equation}
instead of eq.~(\ref{UVdef}). We now impose that $\chi^a_\pm$ is
single-valued at the origin implying that $q^a \pm l\in \ZZ
+\frac{1}{2}$. This implies that $l\mp \eta \in \ZZ$.

Near the origin $U^a$ and $V^a$ satisfy eqs. (\ref{U0}) and
(\ref{V0}), although with a restricted choice of $a$. The leading
order behaviour of the solutions is given by eq.~(\ref{UV0}), and so
the total number of normalisable solutions there is
\begin{equation}
N_0^\pm(l)=\sum_{a=1}^{n_1} I\left(l\ge \mp q^{(1)}_a
+\tfrac{1}{2}\right) + \sum_{a=1}^{n_2} I\left(l\le \pm q^{(2)}_a
-\tfrac{1}{2}\right) \, .
\label{N0d}
\end{equation}
The behaviour of the
spinors at infinity is determined by eqs. (\ref{Ui}) and (\ref{Vi}).
We define $n_{z1}$ and $n_{z2}$ to be the number of massless
fermions with respectively $a\le n_1$ and $a > n_1$. The number of
massive fermions is then $2\bar n$, where $\bar n=n_1-n_{z1} = n_2-n_{z2}$. At
large $r$ the approximate solutions are either exponential or are
given by eq.~(\ref{UVi}). The total number of normalisable
$\chi^a_\pm$ solutions is
\begin{equation}
N^\pm_\infty(l) = \bar n + n_{z1} I\left(l < -\tfrac{C_1}{2}\right) +
n_{z2} I\left(l > \tfrac{C_1}{2} \right) \, .
\end{equation}
Using the fact that $l\mp \eta \in \ZZ $, we can express
$N^\pm_\infty(l)$ in a similar form to eq.~(\ref{N0d}) by
introducing
\begin{equation}
\tilde q^{(1)}_\pm = \pm \frac{1}{2} - \eta
 \mp \left[\frac{C_1}{2} \mp \eta\right]
\label{qtdef1}
\end{equation}
and
\begin{equation}
\tilde q^{(2)}_\pm = \pm \frac{1}{2} + \eta
 \mp \left[\frac{C_1}{2} \pm \eta\right]
\label{qtdef2}
\end{equation}
where $[x]$ is the lowest integer which is strictly greater than
$x$. If $2\eta \in \ZZ$, then the above expressions agree with
eq.~(\ref{qtdef}), and $\tilde q^{(1)}_\pm = \tilde q^{(2)}_\pm $.

We can now write
\begin{equation}
N^\pm_\infty(l) = \bar n + n_{z2} I\left(l\ge \mp \tilde q^{(1)}_\pm +
\tfrac{1}{2}\right) +n_{z1} I\left(l\le \pm \tilde q^{(2)}_\pm -
\tfrac{1}{2}\right) \, ,
\end{equation}
and proceed in a similar way to the previous subsection. We define
$\hat q^{(1)}_a$ to be the combination of the $q^{(1)}_a$ and
$n_{z2}$ copies of $\tilde q^{(1)}_\pm$, and order them so that
$\hat q^{(1)}_1\le \dots \hat q^{(1)}_{\bar n+n_z}$, where
$n_z= n_{z1}+n_{z2}$. Similarly we define $\hat q^{(2)}$ to be the
$q^{(2)}_a$ and $n_{z1}$ copies of $\tilde q^{(2)}_\pm$, and order
them so that $\hat q^{(2)}_1\ge \dots \hat q^{(2)}_{\bar n+n_z}$. This
allows us to write
\begin{eqnarray}
&& \hspace{-0.1in} N_0^\pm(l)+N_\infty^\pm(l)= n_f
\nonumber \\ && \hspace{0.15in} {}
+ \sum_{a=1}^{\bar n+n_z} \left\{
I\left(\mp \hat q^{(1)}_a+\tfrac{1}{2}\le l
\le \pm \hat q^{(2)}_a-\tfrac{1}{2}\right)
- I\left(\pm\hat q^{(2)}_a-\tfrac{1}{2} < l
< \mp\hat q^{(1)}_a+\tfrac{1}{2}\right) \right\} \, . \hspace{0.5in}
\label{Nl2d}
\end{eqnarray}
In contrast to the earlier case, solutions for different $l$ are
independent, and so the total number of real solutions is
$N^\pm = 2\sum_l N^\pm(l)$. Again only half the terms in the above
expression~(\ref{Nl2d}) are non-zero, which allows us to reorder
the summations in the expressions for $N^\pm$, and obtain
\begin{equation}
N^\pm= 2 \sum_{a=1}^{\bar n+n_z}
\left[\pm\hat q^{(1)}_a \pm \hat q^{(2)}_a\right]_+ \, .
\label{Nd}
\end{equation}
Note that there are different definitions of the $\tilde
q^{(i)}_\pm$ in the $\hat q^{(i)}_a$ for the two chiralities. This
last expression is particularly useful in supersymmetry where most
models have block off-diagonal mass matrices.

It is interesting to note that the values of $\tilde q^{(1,2)}_\pm$
depend on the string deficit angle, and hence may change if the string
mass per unit length changes (as may occur at a phase transition).
If the mass matrix has zero eigenvalues, then the number of zero modes
(which will depend on $\tilde q^{(1,2)}_\pm$) could also be altered, and so
$N^\pm$ will be sensitive to changes in the string's gravity. This is
not the case if all the fermion fields are massive away from the string.

\subsection{Zero mass matrix}

Finally, we will determine the number of zero modes when the mass
matrix is zero. In fact $M_{ab} \equiv 0$ is a special case of a
block off-diagonal matrix, although we will cover it separately. In
this case the only things determining whether there are fermion zero
modes are the gauge fields and gravity. We take $\chi^a_\pm =
e^{i(q_a \mp l)\theta} V_a^{\pm *}(r)/\sqrt{C}$. The complex
solution will be normalisable at $r=0$ if $l \leq \pm q_a -1/2$. At
$r=\infty$ it is normalisable if $l> C_1/2$, or equivalently if $l
\geq \mp \tilde q_\pm^{(1)} +1/2$, with $\tilde q^{(1)}_\pm$ given
by eq.~(\ref{qtdef1}). Hence
\begin{equation}
N^+ = 2\sum_{a=1}^{n_f} \left[ q_a +\tilde q_+^{(1)}\right]_+ \, ,
\qquad N^- = 2\sum_{a=1}^{n_f} \left[ -q_a -\tilde q_-^{(1)}\right]_+
\, .
\label{Nnm}
\end{equation}
Let us apply this result to the case treated in ref.~\cite{Stern}.
When $l\in \ZZ$, the compensating factor $\eta=0$, and there is only one
charge $q_a=Q>0$, the number of zero modes is given by
\begin{equation}
N^+= 2\left[Q-\tfrac{1}{2}\right]_+ \, , \qquad N^-=0 \, .
\end{equation}
This is exactly equal to the number of integers satisfying
$0\le m< Q-1$ when $Q\in \ZZ+1/2$.

\section{Supersymmetry examples}
\label{sec:susyex}

In global supersymmetry, one can find zero modes around D- and F-term
strings. As an application of the index theorem, we review some known
examples and count the associated zero modes. We also
consider the case where an F-term string couples to moduli fields as
might be the case in string theory.

\subsection{D-term strings}
\label{ssec:Dterm}

This model uses one superfield $\phi$ which is
charged under an Abelian gauge group. The presence of a
Fayet--Iliopoulos term $\xi$ implies that the potential depends
only on
\begin{equation}
D= -|\phi|^2 +\xi \, .
\end{equation}
The cosmic string is such that $\phi$ interpolates between $\phi=0$
and $\phi=\sqrt \xi$. Other fields may be present, but they will be
zero everywhere for the string solution. The fermion mass matrix in
global SUSY is off-diagonal and involves only $\chi$ and the gaugino
$\lambda$. The winding numbers for the index theorem are
$q^{(1)}_\chi=n$ and $q^{(2)}_\lambda=0$. In supersymmetry, the number
of zero modes is therefore given by~\cite{susyCS}
\begin{equation}
N^+=2n \, , \qquad N^-= 0
\label{DstrN}
\end{equation}
for $n>0$. For negative $n$ the two chiralities are interchanged, and
there are $2|n|$ zero modes with negative chirality.

Using the analysis of the previous section, we can obtain the
approximate form of the zero mode solutions. Taking $n>0$, and
defining $m=-l-1/2$, we find that near the origin the
positive chirality zero modes have the form
\begin{equation}
\lambda \sim r^{m} e^{i(m+1/2)\theta} \, , \qquad
\chi \sim r^{n-1-m} e^{i(n-m-1/2)\theta} \, . \label{mm}
\end{equation}
For the solution to be normalisable, the integer $m$ must satisfy
$0 \leq m \leq n-1$. These approximate solutions are obtained from
eq.~(\ref{UV0}). For large $r$, $\lambda$ and $\chi$ decay exponentially.
As was shown in ref.~\cite{susyCS}, it is also possible to obtain the above
$m=0$ solution by applying a supersymmetry transformation to the
string solution.

\subsection{F-term strings}
\label{ssec:Fterm}

Consider the supersymmetric model with superpotential
\begin{equation}
W=\phi (\phi^+\phi^- -x^2)
\end{equation}
and a $U(1)$ gauge group. $x$ sets the scale of the $U(1)$ symmetry
breaking. The fields $\phi^\pm$ have charges $\pm 1$,
and $\phi$ is uncharged. This model can be used to give hybrid
inflation, with $\phi$ as the inflaton. In that case inflation ends
with the breaking of the $U(1)$ symmetry, which results in the
production of cosmic strings. These have the form
\begin{equation}
\phi^\pm= xf(r) e^{\pm in\theta} \, , \qquad\phi=0 \, .
\end{equation}
The mass matrix of the fermions contains the Yukawa and
gauge interactions
\begin{equation}
-\left(\phi^- \chi - i\sqrt {2} \bar \phi^+ \lambda \right) \chi^+
- \left(\phi^+ \chi + i \sqrt{2}\bar \phi^- \lambda \right)\chi^-
-\phi \chi^+ \chi^-
\end{equation}
where $\chi^i$ are the partners of the $\phi^i$. Since $\phi =0$ this
mass matrix is block off-diagonal, and so we use the index
theorem~(\ref{Nd}). The winding numbers are $q^{(1)}_{\chi+} =n$,
$q^{(2)}_{\chi-}=-n$ and the other two $q$ are zero. Hence we see
that~\cite{susyCS}
\begin{equation}
N^+=2|n| \, , \qquad N^-=2|n| \, .
\end{equation}
There are $|n|$ Weyl zero modes of each chirality.

If our toy model is extended, for example to include additional moduli fields,
then it is possible that their coupling to $\phi$ will mean that
$\phi$ is no longer zero inside the string. The mass matrix will then
cease to have a block off-diagonal form, and the other index
theorem~(\ref{Np1}) will apply. It tells us that now
\begin{equation}
N^+=0 \, , \qquad N^-=0 \, .
\end{equation}
Hence the zero modes are removed if the $\phi=0$ solution is
destabilised. A similar effect was noted in ref.~\cite{susybreak} for
SUSY breaking.

\section{Fermion zero modes in supergravity}
\label{sec:sugra}

\subsection{Fermions in supergravity}
\label{ssec:Lag}

We will now extend our zero mode analysis to supergravity
theories. This is complicated by the inclusion of the gravitino
$\psi_\mu$, whose mass and kinetic terms have a different form to
those used in section~\ref{sec:ind}. We keep the discussion general
and look at cosmic strings with an arbitrary number of chiral
superfields $(\phi^i,\chi^i)$, and one $U(1)$ gauge superfield
$(A_\mu, \lambda)$. Working in Planck units, the fermion part
of the supergravity Lagrangian is
\begin{eqnarray}
{\cal L} &=& \epsilon^{\mu\nu\rho\lambda}
\bar \psi_\mu \bar \sigma_\nu D_\rho \psi_\lambda
-i\bar \lambda \bar \sigma^\mu D_\mu \lambda -iK_{i \bar \jmath} \bar
\chi^{\bar\jmath} \bar \sigma^\mu D_\mu \chi^i \nonumber \\ &&
+\frac{1}{2} \bar \psi_\mu
(D + i\bar\sigma^{\rho\lambda} F_{\rho\lambda})\bar \sigma^\mu \lambda
-\frac{1}{\sqrt 2} K_{i\bar \jmath } D_\nu \phi^i \bar \psi_\mu
\bar \sigma^\nu \sigma^\mu \bar \chi^{\bar \jmath} + i\sqrt{2}K_{i\bar
\jmath } Q \phi^i \bar \chi^{\bar \jmath} \bar \lambda \nonumber
\\ && - m_{3/2} \bar \psi_\mu \bar \sigma^{\mu\nu} \bar \psi_\nu +
\frac{i}{\sqrt{2}} m_i \bar \psi_\mu \bar \sigma^\mu \chi^i -
\frac{1}{2} m_{ij} \chi^i \chi^j + \Lambda \sigma^\mu \bar \psi_\mu
+ (\mathrm{c.c.})
\label{Lf}
\end{eqnarray}
where $\bar \sigma^{\mu \nu}
= (\bar \sigma^\mu \sigma^\nu - \bar \sigma^\nu \sigma^\mu)/4$.
We have defined $m_{3/2}= e^{K/2}W$, $m_i= e^{K/2} D_i W$ and
$m_{ij}= e^{K/2} D_i D_j W$ with $D_i W= \partial_i W + K_i W$.

In supergravity, the supersymmetry transformations preserve the
Lagrangian, and act as gauge transformations. When analysing the
theory it is important to fix the gauge. We do this by imposing
\begin{equation}
\bar \sigma^\mu \psi_\mu=0 \, .
\label{psigauge}
\end{equation}
To achieve this we have incorporated a Lagrange multiplier
$\Lambda$, in the above Lagrangian. In general this does not fix the
gauge symmetry fully. Residual gauge symmetries are still present
corresponding to $\bar \sigma^\mu\delta \psi_\mu=0$. In the
following we will not fix them.

For the above theory, the fermion supersymmetry transformations are
\begin{equation}
\delta \chi^i= i\sqrt 2 \sigma ^\mu \bar \epsilon D_\mu \phi^i
-\sqrt 2 e^{K/2}{K^{i\bar \jmath}} D_{\bar \jmath} \bar W \epsilon
\end{equation}
\begin{equation}
\delta \lambda = (F_{\mu\nu} \sigma^{\mu\nu} - i D) \epsilon \, .
\end{equation}
The gravitino variation is
\begin{equation}
\delta\psi_\mu= 2 \left(\partial_\mu + \omega_\mu + \frac{i}{2}
A^B_\mu\right) \epsilon +im_{3/2} \sigma_\mu \bar \epsilon
\label{graviti}
\end{equation}
where $A^B_\mu$ is defined in eq.~(\ref{AB}).

In general, cosmic string solutions break all supersymmetries. However
there exist BPS cosmic strings which preserve
half of the original supersymmetries. These configurations are
solutions of the Killing spinor equations obtained by equating some of
the above fermion variations to zero.
BPS configurations are of particular interest in string theory.

Let us now look for solutions of the Killing spinor equations in a
cosmic string background, where the fields depend on $r$ and $\theta$
only. Such solutions will be BPS cosmic strings. The $t$ and $z$
components of the gravitino variation lead to the equations
\begin{equation}
B' \sigma^1 \epsilon =  i m_{3/2} \bar \epsilon \, , \qquad 
B' \sigma^1 \epsilon = -i m_{3/2} \bar \epsilon
\end{equation}
whose only solution is obtained when both $m_{3/2}=0$ and $B'=0$.
Hence a non-zero gravitino mass is not compatible with the existence
of BPS states. The conformal factor $e^{2B}$ must also be a
constant.

Taking $\epsilon(r,\theta) = e^{i\sigma^3 \theta/2} \epsilon_0(r)$,
the variations of the other fields for a cosmic string background are
\begin{equation}
\delta \lambda = -i(F_{12} \sigma^3 +D)\epsilon
\label{cstrf1}
\end{equation}
\begin{equation}
\delta\chi^i = \sqrt{2}i\left(
\sigma^r \partial_r \phi^i +\sigma^\theta D_\theta \phi^i\right) \bar \epsilon
\end{equation}
\begin{equation}
\delta \psi_\theta = i \left(\sigma^3 -2iw_\theta + A^B_\theta\right) \epsilon
\, , \qquad\delta \psi_r = 2 \partial_r \epsilon \, .
\label{cstrf3}
\end{equation}
For a BPS cosmic string, these must all vanish for some choice of
$\epsilon$. This condition gives the required field equations for the
string solution. We take $\epsilon_0(r)$ to be a constant and an eigenstate
of $\sigma^3$. From this we obtain the BPS
equations~(\ref{Dstr1})--(\ref{Dstr3}). Note that they are first order
differential equations, and that 1/2 of the supersymmetry is preserved.

\subsection{Gravitino field equations}

Using the previous Lagrangian, the gravitino equations are given by
\begin{eqnarray}
&&\epsilon^{\mu\nu\rho\lambda} \bar \sigma_\nu D_\rho \psi_\lambda
+ \frac{1}{2} (D+ i F_{\rho \lambda}\bar \sigma^{\rho \lambda})
\bar \sigma^\mu \lambda
-\frac{1}{\sqrt 2} K_{i\bar \jmath } D_\nu \phi^i \bar \sigma^\nu
\sigma^\mu \bar \chi^{\bar \jmath}
\hspace{1.5in} \nonumber \\ && \hspace{2.5in} {} 
-2 m_{3/2} \bar \sigma^{\mu\nu} \bar \psi_\nu
+\frac{i}{\sqrt{2}} m_i \bar \sigma^\mu \chi^i =\bar\sigma^\mu\Lambda
\end{eqnarray}
and the spin 1/2 equations are
\begin{equation}
-i\bar \sigma^\mu D_\mu \lambda + i\sqrt 2 K_{i\bar \jmath} T^s \phi^i
\bar \chi^{\bar \jmath} + \frac{1}{2}\bar \sigma^\mu (D +i
F_{\rho \lambda}\sigma^{\rho \lambda})\psi_\mu =0
\end{equation}
\begin{equation}
-i \bar \sigma^\mu D_\mu \chi^i +i\sqrt 2 T^s \phi^i \bar \lambda -
\frac{1}{\sqrt 2} D_\nu \phi^i \bar \sigma^\mu \sigma^\nu \bar \psi_\mu -
K^{\bar \jmath i} \bar m_{\bar \jmath \bar k} \bar \chi^{\bar k}
-\frac{i}{\sqrt 2} K^{\bar \jmath i} \bar m_{\bar \jmath} \bar
\sigma^\mu \psi_\mu=0 \, .
\end{equation}
This set of equations can be mapped to the Dirac equations used in
the proof of the index theorem. No manipulation is necessary for the
spin 1/2 fermions. For the gravitino, more work is needed.

The fact that our cosmic string background is
independent of $z$ and $t$, simplifies the above equations. In
particular we see that the connection terms in the covariant
derivatives $D_{z,t} = \partial_{z,t} + w_{z,t}$ are
\begin{equation}
 w_z = i\sigma^2B'/2 \, , \qquad w_t = -\sigma^1 B'/2
\end{equation}
which depend only on $B'$. We also have $w_r =0$. The
$D_\theta$ derivatives are
\begin{eqnarray}
&& D_\theta \bar \psi_\mu= \left(\partial_\theta-\frac{i\sigma^3}{2}C' -
\frac{i}{2}A_\theta^B\right)\bar \psi_\mu \, , \qquad D_\theta \lambda=
\left(\partial_\theta-\frac{i\sigma^3}{2}C' +
\frac{i}{2}A_\theta^B\right)\lambda \nonumber \\ && D_\theta \chi^i=
\left(\partial_\theta-\frac{i\sigma^3}{2}C' -iQ_i A_\theta -
\frac{i}{2}A_\theta^B\right)\chi^i + \Gamma^i_{jk} D_\theta \phi^j \chi^k
\end{eqnarray}
where $A_\mu^B$ is given by eq.~(\ref{AB}).

After eliminating the Lagrange multiplier, one obtains three
equations which can be cast in Dirac-like form. After gauge fixing,
the gravitino field has 3 components. For a cosmic string background
it is convenient to write them in terms of the three independent
Weyl fermions
\begin{equation}
\Sigma= \sigma^r \bar \psi_r +\sigma^\theta \bar \psi_\theta
\, , \qquad
\Psi=\sigma^r \bar \psi_r - \sigma^\theta \bar \psi_\theta
\, , \qquad
\Pi= \sigma^t \bar \psi_t - \sigma^z \bar\psi_z \, .
\label{PSP}
\end{equation}
The gravitino equations can then be expressed as
\begin{equation}
i(\sigma^r [\partial_r +B'] + \sigma^\theta D_\theta) \Pi
+ 2 i \sigma^z w_z \Psi - m_{3/2} \bar \Pi
-i(\bar \sigma^t\partial_t-\bar \sigma^z\partial_z)\Sigma = 0
\label{Pieq}
\end{equation}
for the $\Pi$ equation, and
\begin{equation}
i(\sigma^r [\partial_r +B'] - \sigma^\theta D_\theta) \Sigma
-\sqrt{2} K_{\bar \imath j} [\sigma^r\partial_r -
\sigma^\theta D_\theta] \bar \phi^{\bar \imath} \chi^{j}
 + m_{3/2}\bar \Psi
+i(\bar \sigma^t\partial_t+\bar \sigma^z\partial_z)\Psi = 0
\label{Sigmaeq}
\end{equation}
for the $\Sigma$ equation and a combination
\begin{eqnarray}
&& i[\sigma^r (\partial_r +C'/C) - \sigma^\theta \tilde D_\theta]\Psi
-i[\sigma^r (\partial_r +C'/C) + \sigma^\theta \tilde D_\theta]\Sigma
 +2i\sigma^z w_z \Pi +2 \sigma^3 F_{12} \bar \lambda
\nonumber \\ && {}
-\sqrt{2}K_{\bar \imath j}
\sigma^\nu D_\nu \bar \phi^{\bar \imath} \chi^j + 2m_{3/2}\bar \Sigma
+i(\bar \sigma^t\partial_t-\bar \sigma^z\partial_z)\Pi
+i(\bar \sigma^t\partial_t+\bar \sigma^z\partial_z)\Sigma
=0
\label{Psieq}
\end{eqnarray}
closing the system. We have defined $\tilde D_\theta
=\partial_\theta +i \sigma^3 C'/2 -iA^B_\theta/2$. These equations
are coupled in general.

In a similar manner the spin 1/2 equations are given by
\begin{equation}
i(\sigma^r \partial_r + \sigma^\theta D_\theta) \lambda + i\sqrt 2
K_{i\bar \jmath} T^s \phi^i \bar \chi^{\bar \jmath} - \sigma^3
F_{12} \bar \Sigma -i(\bar \sigma^t\partial_t+\bar \sigma^z
\partial_z) \lambda = 0 \label{lam}
\end{equation}
and
\begin{eqnarray}
&& i(\sigma^r D_r + \sigma^\theta D_\theta)\chi^i +i\sqrt 2
T^s \phi^i \bar \lambda - K^{\bar \jmath i} \bar m_{\bar \jmath \bar
k} \bar \chi^{\bar k} \nonumber \\ && {} + \frac{1}{\sqrt 2}
\sigma^\nu D_\nu \phi^i \Sigma + \frac{1}{\sqrt 2}[\sigma^r
\partial_r -\sigma^\theta D_\theta]\phi^i \Psi - i(\bar
\sigma^t\partial_t+\bar \sigma^z \partial_z) \chi^i =0  \, . \label{chi}
\end{eqnarray}
The zero mode solutions we are looking for depend on $r$ and
$\theta$, and so we can ignore the $z$ and $t$ derivatives. For
fermion currents on the string (which are physically more
interesting), the $z$ and $t$ dependence of the fields will have to be
included. We will discuss this in section~\ref{sec:curr}.

Let us now discuss the normalisation of the gravitino fields. We choose
a norm compatible with the requirements of canonical quantisation in
field theory
\begin{equation}
||\psi_\mu||^2 = \frac{i}{2} \int dr d\theta e^{2B} C \left[ \Sigma
P_\Sigma + \Pi P_\Pi + \Psi P_\Psi -(\mathrm{c.c.})\right] \, .
\end{equation}
We identify $P_X= \delta {\cal L}/(\delta \partial_0 X)$ as the
conjugate momentum, where ${\cal L}$ is the Lagrangian~(\ref{Lf}). The
above norm for the gravitino part of the wavefunction is equal to
\begin{equation}
||\psi_\mu||^2 = \frac{1}{2} \int dr d\theta C e^{2B} \bar \psi_\mu
\eta^{\mu \nu}\psi_\nu \, .
\end{equation}
The full norm for the fermion states is then
\begin{equation}
\int dr d\theta C e^{2B} \left[ \sum_i |\chi^i|^2 + \frac{1}{2}\left(
|\Psi|^2 + |\Sigma+\Pi|^2 -|\Pi|^2 \right) \right] \, .
\label{gnorm}
\end{equation}
This must be finite if the zero mode solution is to be a
normalisable bound state. Note that the positivity of the norm is
not guaranteed. However since we expect the string to be stable, any
normalisable fermion bound states should have non-negative norms.
It is still possible that they are gauge artifacts, and so there is
a risk that our analysis leads to some unphysical zero modes. If we
had been working in Minkowski space (instead of a string background)
we could have removed any unphysical modes with further gauge
fixing. It is not at all clear how to do this in a string background
and we leave the possibility of using residual supersymmetries to
further work.

\subsection{Index theorem}

We see that the Dirac-like equations for the gravitino
fields~(\ref{Pieq})--(\ref{Psieq}) have a
different form to that assumed in eqs.~(\ref{dirind1}) and (\ref{dirind2}),
which were used to derive the number of fermion zero modes. The
expressions derived in section~\ref{sec:ind} cannot therefore be
applied directly to models which include gravitinos. However, the
number zero modes can still be determined by analysing approximate
large and small $r$ solutions. As we will show, this does allow our
index theorem to be modified to include gravitinos.

We will start by extending the index theorem for block off-diagonal mass
matrices~(\ref{Nd}), as this will be the relevant case for
D-strings. In this case the gravitino mass term $m_{3/2}$ must be
zero. For large and small $r$ the gravitino Dirac equations reduce to
\begin{equation}
\left(\partial_r \mp \frac{i}{C}\left[\partial_\theta
\mp \frac{i}{2}C'+i q_\psi a \right]\right)\Sigma_\pm = 0
\label{Sigmaapp}
\end{equation}
\begin{equation}
\left(\partial_r +\frac{C'}{C} \mp \frac{i}{C}\left[\partial_\theta
\pm \frac{i}{2}C'+i q_\psi a \right]\right) \Psi_\pm
\propto \ \frac{i}{C}\partial_\theta \Sigma_\pm \, , \ B' \Pi_\pm
\label{Psiapp}
\end{equation}
\begin{equation}
\left(\partial_r \pm \frac{i}{C}\left[\partial_\theta
\mp \frac{i}{2}C'+i q_\psi a \right]\right)\Pi_\pm \propto B' \Psi_\pm
\label{Piapp}
\end{equation}
where we have defined $q_\psi = -n\xi/2$.
We have neglected $A^B_\mu - \xi A_\mu$, $D_\mu\phi^i$,
and $\sigma^3 F_{12}$ [which would form part of $M_{ab}$ and
$J_\mu$ in eq.~(\ref{Dchi})], since they are all subdominant as $r$
tends to 0 or $\infty$.

These equations are analogous to a Dirac equation and can be analysed
both at the origin and at infinity. Following the ansatz~(\ref{UVd}) we take
\begin{equation}
\Psi=\frac{V^\Psi_\pm}{\sqrt C}e^{ -i(q_\psi \mp l) \theta}
\, , \qquad
\Sigma=\frac{V^\Sigma_\pm}{\sqrt C}e^{ -i(q_\psi \mp l) \theta}
\, , \qquad
\Pi = \frac{U^\Pi_\pm}{\sqrt C}e^{ i(-q_\psi \pm l) \theta} \, .
\end{equation}

As in section~\ref{sec:ind}, we look for the leading order
behaviour of the fermion equations near the origin. The gauge field
$A_\theta$ is negligible there, so we can ignore it in
eqs.~(\ref{Sigmaapp})--(\ref{Piapp}).

We start with a solution whose leading order behaviour comes from
$\Sigma$. This is obtained by solving eq.~(\ref{Sigmaapp}).
Eq.~(\ref{Psiapp}) implies $\Psi$ will have similar behaviour. The
behaviour of other fields comes from their coupling to $\Sigma$ and $\Psi$ via
Yukawa terms. Since the string solution is regular at the origin, we
find the other fermion fields are at least as well behaved as $\Sigma$
there. Putting this all together gives the leading order behaviour of
the solution
\begin{equation}
V^\Sigma_\pm, V^\Psi_\pm \sim r^{\pm q_\psi+1 -l}
\, , \qquad \chi^i,\lambda,\Pi = O(r^{\pm q_\psi-l+2}) \, .
\end{equation}
Another solution is obtained by taking the right-hand side of
eq.~(\ref{Psiapp}) to be subdominant, and then solving for $\Psi$. The
leading behaviour of the other fields then comes from their coupling
to $\Psi$ via Yukawa terms. Hence
\begin{equation}
V^\Psi_\pm \sim r^{\pm q_\psi-1 -l}
\, , \qquad
\chi^i,\lambda,\Sigma,\Pi = O(r^{\pm q_\psi-l}) \, .
\end{equation}
Finally, we take the right-hand side of eq.~(\ref{Piapp}) to be
subdominant, and solve for $\Pi$ to obtain the third independent
solution. Substituting this expression for $\Pi$ into
eq.~(\ref{Psiapp}), and using $B' \sim r$ near the origin, we can
obtain the leading behaviour of $\Psi$ and $\Sigma$. As with the other
two solutions, the behaviour of the non-gravitino fields is determined
by their couplings to Yukawa terms. We find
\begin{equation}
U^\Pi_\pm \sim r^{\mp q_\psi+l}
\, , \qquad
V^\Sigma_\pm, V^\Psi_\pm \sim r^{\mp q_\psi + l+2}
\, , \qquad \chi^i,\lambda = O(r^{\mp q_\psi+l+3}) \, .
\end{equation}
Hence we need respectively $l \leq \pm q_\psi +1/2$, 
$l \leq \pm q_\psi -3/2$ and $l \geq \mp q_\psi- 1/2$ for each of the
above three solutions to be normalisable inside the string core. In
order to use the same algebra that leads to the expression~(\ref{Nd}),
we want to write the above conditions in the same form as
eq.~(\ref{N0d}). This is achieved by introducing the chirality
dependent effective charges
\begin{equation}
q_\Psi^{(2)} = q_\psi \mp 1 \, , \qquad
q_\Sigma^{(2)} = q_\psi \pm 1 \, , \qquad
q_\Pi^{(1)} = -q_\psi \pm 1 \, .
\label{qgrav}
\end{equation}
Notice the shifts of one unit.

We also need to consider the behaviour of the other fermion solutions,
whose leading behaviour comes from $\chi^i$ and $\lambda$. For these
solutions the gravitino fields are subdominant, and so the analysis of
section~\ref{sec:ind} still applies.

The gravitino is massless at infinity and so, as was discussed in
section~\ref{sec:ind}, some zero modes with low angular momentum will
be lost. We can again account for this by introducing some effective
winding numbers $\tilde q_\pm$. For the gravitino fields we find the
leading order behaviour of three approximate solutions at infinity is
\begin{equation}
V^\Psi_\pm \sim r^{-1 -l/C_1}
\end{equation}
\begin{equation}
V^\Sigma_\pm \sim V^\Psi_\pm \sim r^{1 - l/C_1}
\end{equation}
\begin{equation}
U^\Pi_\pm \sim r^{l/C_1} \, , \qquad V^\Sigma_\pm, V^\Psi_\pm \sim r^{l/C_1}
\end{equation}
where we have used $B' \sim 1/r$ far away from the string.
Following the reasoning behind eq.~(\ref{qtdef1}), we see that the
above solutions are normalisable if
$l\geq \mp\tilde q^{(1) }_{\Psi\pm} +1/2$,
$l\geq \mp\tilde q^{(1) }_{\Sigma\pm} +1/2$ and
$l\leq \pm\tilde q^{(2) }_{\pm} -1/2$ with
\begin{equation}
\tilde q^{(1) }_{\Psi\pm} = \pm \frac{1}{2} -\eta \mp \left[ -\frac{C_1}{2}
\mp \eta\right]
\label{qtpsi}
\end{equation}
\begin{equation}
\tilde q^{(1) }_{\Sigma\pm} = \pm \frac{1}{2} -\eta \mp \left[\frac{3C_1}{2}
\mp \eta\right]
\label{qtsigma}
\end{equation}
and $\tilde q^{(2) }_{\pm}$ is defined in eq.~(\ref{qtdef1}).
The single-valuedness of the gravitino implies that we take
$\eta= (1 - n\xi)/2$. Using $0 < C_1 \leq 1 -|n|\xi$, we find that
\begin{equation}
\tilde q^{(1) }_{\Psi\pm} = \frac{n\xi}{2} = -q_\psi
\end{equation}
\begin{equation}
\tilde q^{(1) }_{\Sigma\pm}
= \frac{n\xi}{2} \mp I\left(3[1-C_1] \leq 2 \pm n\xi\right)
\end{equation}
\begin{equation}
\tilde q^{(2) }_{\pm}
= -\frac{n\xi}{2} \mp I\left(1-C_1 = \mp n\xi\right)
\end{equation}
hence $\tilde q^{(1) }_{\Sigma\pm} = -q^{(2)}_\Sigma$ unless the strings have
a large deficit angle (which implies they must be very heavy). For
non-BPS strings, we always have $\tilde q^{(2) }_{\pm} = q_\psi$.

With the aid of the modified charges $q^{(2)}_{\Sigma,\Psi}$ and
$\tilde q^{(1) }_{\Psi,\Sigma \, \pm}$, we can now use the index
theorem~(\ref{Nd}) to obtain the total number of zero modes, even when
gravitinos are included in the model.

If the fermion mass matrix is more generic, i.e.\ is not block
off-diagonal, we must instead use the ansatz
\begin{equation}
\Psi= \frac{e^{-iq_\psi \theta}}{\sqrt C} \left\{
U^{\Psi*}_{\pm} e^{\mp i l\theta} + V^\Psi_{\pm}e^{\pm il\theta}\right\}
\end{equation}
\begin{equation}
\Sigma= \frac{e^{-iq_\psi \theta}}{\sqrt C} \left\{
U^{\Sigma*}_{\pm} e^{\mp i l\theta} + V^\Sigma_{\pm}e^{\pm il\theta}\right\}
\end{equation}
\begin{equation}
\Pi= \frac{e^{-iq_\psi \theta}}{\sqrt C} \left\{
U^{\Pi}_{\pm} e^{\pm i l\theta} + V^{\Pi*}_{\pm}e^{\mp il\theta}\right\} \, .
\end{equation}
The single-valuedness of the wave functions implies that $q_\psi$ must
be integer or half integer.

As before we are interested in the leading order behaviour of the
above solutions at the origin. Using the effective charges defined
above, it is given by eq.~(\ref{UV0}). If $m_{3/2} \neq 0$, then the
solutions will have exponential behaviour at infinity, and the
analysis of subsection~\ref{ssec:in1} will apply. If on the other hand
$m_{3/2} =0$, then some of the large $r$ gravitino solutions will have
power law behaviour. As before, we can deal with this by including
some extra $\tilde q_\pm$ winding numbers. Instead of the
expression~(\ref{qtdef}), we use the above effective winding
numbers~(\ref{qtpsi})--(\ref{qtsigma}). With all these modifications, the
index theorem~(\ref{Np1}) will now apply to models with gravitinos.

The expressions~(\ref{Nd}) and (\ref{Np1}) now give the number of
fermion bound states with finite norm. However if the field $\Pi$ is
non-zero we cannot be certain that the norm is positive, and so some
of these solutions may be gauge artifacts or even ghosts. If the
wavefunctions are dominated by the other fields, the norm will be
positive, so it is only solutions whose leading order behaviour
comes from $\Pi$ which are likely to be gauge artifacts. This
suggests that we may be able to avoid counting the gauge zero modes
by excluding the winding numbers corresponding to $\Pi$ from the
analysis. However this is not guaranteed to work, and so we will
consider both possibilities in what follows.

\section{Supergravity examples}
\label{sec:sugex}

We will now apply our index theorem to cosmic strings in supergravity
models. In contrast to the models discussed in
section~\ref{sec:susyex}, we must now include the gravitino
field.

\subsection{D-strings}
\label{ssec:Dstr}

For the global SUSY equivalent of the D-string (see
subsection~\ref{ssec:Dterm}), it was possible to find a fermion bound
state using the broken supersymmetry transformation. The same idea can
be tried for D-strings. The transformations for a string background
are given by eqs.~(\ref{cstrf1})--(\ref{cstrf3}).
For $n>0$, the Killing spinor $e^{-i\theta/2} \epsilon_{0-}$ corresponds to the
preserved 1/2 supersymmetry. In supergravity, the spinor
$e^{i\theta/2} \epsilon_{0+}(r)$ is the Goldstino of broken
supersymmetry obtained in ref.~\cite{rachel}. In order to satisfy the
gauge choice~(\ref{psigauge}) we need
$\epsilon_0(r) \propto \exp[\int (1-C')/C dr]$. From
eq.~(\ref{cstrf3}) we obtain
\begin{equation}
\bar \Psi = \bar \sigma^r \psi_r-\bar \sigma^\theta \psi_\theta
= 4 \frac{1-C'}{C} \sigma^1 \epsilon_{0+}(r) \, .
\end{equation}
Hence $\Psi \propto n\xi r^{-2+1/C_1}$ at infinity, and it is
therefore not a normalisable bound state. This contrasts with the
situation for global supersymmetry, where the corresponding state is
localised on the string. The disappearance of this zero mode can be
confirmed using the index theorem.

Since we have a BPS string, the $m_{3/2}=0$ and the $B'$ terms in the
gravitino equations~(\ref{Pieq})--(\ref{Sigmaeq}) vanish. We see that
$\Pi$ decouples from the other fields. It is pure gauge so we can
ignore it.

Taking $n>0$, we see that the field $\Sigma$ also decouples for
positive chirality states. Solving eq.~(\ref{Sigmaeq}) we find $\Sigma=0$.
The fermion mass matrix is block off-diagonal, and so we separate
the fields into two groups. The first
contains the Higgsino with winding number $q^{(1)}_\chi = n -q_\psi$, and the
second contains the gaugino ($q^{(2)}_\lambda = q_\psi$), and the
gravitino field $\Psi$ ($q^{(2)}_{\Psi\pm}= q_\psi \mp 1$). Since the
gravitino field is massless at infinity we need to add
the effective winding $\tilde q^{(1) }_{\Psi \pm}=-q_\psi$.
For the negative chirality states $\Sigma$ does not decouple, and so in
this case so we also need to include $q^{(2)}_{\Sigma-}$ and
$\tilde q^{(1)}_{\Sigma-}$ in the analysis. Using the index
theorem~(\ref{Nd}) we find that (for $n>0$) the number of zero modes is
\begin{equation}
N^+=2(n-1) \, , \qquad N^-=0 \, .
\end{equation}
Comparing this with the results for global SUSY~(\ref{DstrN}),
we see that in the broken $N=1/2$ sector corresponding to the
positive chirality, a pair of zero modes, i.e.\ a complex spinor, has
disappeared. In ref.~\cite{rachel} this result was suggested using a
superHiggs argument, i.e.\ the Goldstino field does not lead to zero
modes anymore as it is eaten by the gravitino. Our calculations
justify this heuristic reasoning. This result is remarkable as it
exemplifies the role of supergravity in the counting of zero modes.

Let us be more explicit and show which zero mode disappears in
supergravity. It is helpful to introduce $m = -l -1/2 +q_\psi$, which
is an integer. Near $r=0$, the general solution to the fermion
field equations has the leading order behaviour
\begin{equation}
U^\chi = c_1 r^{n-m-1/2} \, , \qquad V^\lambda = c_2 r^{m+1/2}
\, , \qquad V^\Psi = c_3 r^{m-1/2}
\end{equation}
where the $c_i$ are constants. At infinity one combination of solutions
decays exponentially, and (for $m\geq 0$) is the only normalisable
solution there. In general, the form of this combination of solutions
near $r=0$ will be given by the above expression with all $c_i \neq 0$.
Hence if the solution is to be normalisable everywhere, we must have
$m \leq n -1$, $m \geq 0$ and $m \geq 1$ (the last condition comes from the
gravitino field). Without the gravitino we would need $n-1 \geq m \geq 0$.
Including it we lose the $m=0$ mode as it cannot be normalisable close
to the origin. The rest of the zero mode tower is preserved.

\subsection{Non-BPS D-strings}

We now suppose that the above strings are no longer BPS and that
$B' \neq 0$. These strings have the same field content as
BPS ones, but $T^r_r$ is no longer zero. We see that the fields $\Pi$ and
$\Sigma$ no longer decouple, and the analysis is changed. If we
exclude the $q$ corresponding to $\Pi$, we obtain almost the same
result as for the BPS D-strings (for $n>0$)
\begin{equation}
N^+=2(n-1) +2 I\left(3[1-C_1] > 2 + n\xi\right) \, , \qquad N^-=0 \, .
\end{equation}
The $I(\ldots)$ term will be zero unless the string deficit angle is
very big. If this is not the case the results of the previous section
hold, although the fermion wavefunction will now include small
contributions from the $\Pi$ and $\Sigma$ fields.

On the other hand if we do include the additional charges
$q_\Pi^{(1)} = -q_\psi \pm 1$ and $\tilde q_\pm^{(2)} = q_\psi$ for
the field $\Pi$, the index $N^+$ is increased by two. These extra zero
modes do satisfy all the requirements for normalisability, although we
cannot be sure that they have positive norm, and they may just be
gauge. The corresponding solutions are gauge for BPS D-strings,
suggesting they will also be gauge in this case, and should be ignored.

If the strings are very heavy, and $3(1-C_1) > 2 + n\xi$, there will
be an additional $m=-1$ zero mode. Near the origin the leading order
behaviour of the solutions is
\begin{equation}
U^\chi = c_1 r^{n-m-1/2} \, , \qquad V^\Psi \sim V^\Sigma = c_2 r^{m+3/2}
\, , \qquad
U^\Pi = c_3 r^{-m-1/2}
\end{equation}
and near infinity it is
\begin{equation}
V^\lambda \sim U^\chi = c_4 e^{-\sqrt{2} \xi} \, , \quad
V^\Sigma \sim r^{(m+1/2-q_\psi)/C_1 + 1} \, , \quad
V^\Psi = c_5 r^{(m+1/2-q_\psi)/C_1 - 1} + c_6 V^\Sigma \, .
\end{equation}
Hence if $m=-1$ and $3C_1 < 1-n\xi$, there is a solution to the
fermion field equations which has the above form and is normalisable
everywhere. We see here that the fermion wavefunctions have power law
decay outside the string. However the above normalisable solution is
only possible if the string deficit angle is very big, and such heavy
strings are ruled out by astrophysical observations.

\subsection{D-strings with spectators}

We add to the D-strings a new field $\Phi$ and a superpotential
\begin{equation}
W= \frac{a}{2} \phi \Phi^2 \, .
\end{equation}
The fermion $\chi_\Phi$ associated with $\Phi$ has winding $q_\Phi=-n/2$ as its
charge is $Q_\Phi= -1/2$. In the string background, we have $\Phi=0$,
and the derivatives $\partial W/\partial \phi^i$ vanish too.

The field $\chi_\Phi$ does not couple to any of the other fermions,
and so it can be analysed independently. Its presence does not affect
the results of the previous subsections. Applying the index
theorem (with $n>0$) we find
\begin{equation}
N^+=0 \, , \qquad N^-=n \, .
\end{equation}
The presence of these chiral zero modes has been obtained in
ref.~\cite{binetruy}. The above result also holds for global SUSY, and
is not affected by the inclusion of SUGRA effects.

\subsection{Massless spectators on D-strings}

It is also possible for fermion fields which do not couple to a
Higgs field to have zero modes on a D-string. In contrast to fermions
which are massive off the string, the number of zero mode solutions is
sensitive to the deficit angle of the string, which itself depends on
whether the string is BPS or not.

Consider a fermion with winding $q=q_\psi + Q n$, where $Q>0$ and
$Qn$ is an integer. The expression~(\ref{Nnm}) will give the number of zero
modes in this case. Taking $n>0$, we have $\tilde q^{(1)}_+ =
-q_\psi-1$ and $\tilde q^{(1)}_- = -q_\psi$ for a BPS string. Hence
there are $nQ-1$ positive chirality zero modes. On the other hand
for a non-BPS string, the deficit angle is increased, which alters
the value of $\tilde q^{(1)}_+$ to $-q_\psi$. This results in an
additional zero mode, and so the conductivity of the string is
actually increased by supersymmetry breaking. If we have $q=-q_\psi + Q n$
instead, the equivalent calculation gives $Qn$ positive chirality zero modes,
irrespective of whether the string is BPS or not.

Note that these zero modes have power law decay outside the string core, in
contrast to the usual exponential decay. This implies corresponding currents
will be less effective at stabilising vortons. The radius of a typical
vorton is about ten times the size of the string core. If the zero
mode wavefunctions decay exponentially outside the string core, the
overlap between the wavefunctions of fermion states on opposite sides of the
string loop will be tiny. On the other hand if the wavefunctions have
power law decay, this overlap will be large, and we expect the fermions
to scatter off each other, producing particles which are not confined
to the string. Hence the current which was stabilising the loop will
decay, and the vorton will collapse. However, we expect that some
loops with very large radii will still be stable, but only a small
number of such vortons will be produced by a string network. The corresponding
vorton constraints are therefore much weaker than usual~\cite{CD}.

\subsection{Non-BPS F-term strings}

For non-BPS strings the three gravitino degrees of freedom $\Psi$,
$\Sigma$ and $\Pi$ do not decouple, and we need to include the
corresponding effective winding numbers defined in
eq.~(\ref{qgrav}). For non-zero $m_{3/2}$ the mass matrix is not
block off-diagonal, and we must use the index theorem for generic
mass matrices~(\ref{Np1}). The winding numbers, $q_a$ of the
fermions must then be all either half integer or integer.

As an example, we extend the F-term string model of
subsection~\ref{ssec:Fterm}. This has no Fayet--Iliopoulos terms,
and the three effective gravitino charges~(\ref{qgrav}) reduce to
$\mp 1$, $\pm 1$ and $\pm 1$. As before the Higgsinos have charges
$n$, $-n$, $0$ and the gaugino is also chargeless. Since we are now
using a generic mass matrix, the different chirality Higgsino zero
modes mix together to form massive states, and the corresponding
zero modes disappear. We discussed a similar effect for the SUSY
F-term model, although in this case the mixing arose from an extra
non-trivial Yukawa term. This could also occur for our SUGRA model
(perhaps due to the effects of other sectors of the theory),
although the gravitino coupling alone is enough to remove the zero
modes. Hence the inclusion of gravitinos means that none of the
global SUSY zero modes survive.

Using the index theorem we find that $N^+= N^-=2$. These extra zero
modes arise from the inclusion of the $\Pi$ gravitino field in the
analysis. The fact that the norm is not positive definite suggests
these may be gauge degrees of freedom. This appeared to be the case
with the corresponding modes for the non-BPS D-strings, and so we
suspect it is also the case here. If we remove the $q$ arising from
$\Pi$, we obtain $N^+=N^-=0$, and so there are no fermion zero
modes.

\section{Fermion currents}
\label{sec:curr}

\subsection{Extending zero modes to massless currents}

The zero mode solutions we have found depend only on $r$ and
$\theta$. We will show that for theories without gravitinos these can
easily be extended to massless currents by adding a $z$ and $t$
dependent phase to the zero mode solution.

We see from the field equations~(\ref{lam}) and (\ref{chi}) that if we ensure 
\begin{equation}
(\bar \sigma^t\partial_t+\bar \sigma^z
\partial_z) \lambda = (\bar \sigma^t\partial_t+\bar \sigma^z
\partial_z) \chi^i = 0
\end{equation}
then the $r$ and $\theta$ dependence of our zero mode solutions will
still satisfy the equations of motion. Let us make the changes 
$U_\pm \to f_\pm(z,t)U_\pm$ and $V_\pm \to f_\pm(z,t)V_\pm$ to the
fermion ansatz~(\ref{UVdef}). Taking $f_\pm(z,t) = e^{i\omega (t \mp z)}$, 
the above conditions hold, and so this extension of our
fermion zero mode solution also satisfies the field equations. We see
that it has definite chirality (like the zero mode), and that it moves
at the speed of light along the string. The direction of the current
is determined by its chirality.

Let us now consider what happens in the presence of
the gravitino field. Extending the above arguments to the zero modes
on a BPS D-string (see subsection~\ref{ssec:Dstr}), we try the ansatz
\begin{equation}
\chi_\pm(x^\mu)= f_\pm(z,t) \chi_\pm (r,\theta) \, , \ \ 
\lambda_\pm(x^\mu)= f^*_\pm(z,t) \lambda_\pm (r,\theta) \, , \ \ 
\Psi_\pm(x^\mu)= f_\pm(z,t) \Psi_\pm (r,\theta) \, .
\label{curr}
\end{equation}
Since $(\bar \sigma^t\partial_t+\bar \sigma^z \partial_z) \Psi =0$,
the ansatz works, and we again have a massless current with definite chirality.

However if we try to include the other gravitino fields $\Sigma$ and
$\Pi$ (as would be required for non-BPS strings),
we find that the above extension of zero modes to currents runs into major
difficulties. First of all we see that both 
$(\bar \sigma^t\partial_t + \bar \sigma^z \partial_z) \Sigma$ and
$(\bar \sigma^t\partial_t - \bar \sigma^z \partial_z) \Sigma$ appear
in the gravitino equations~(\ref{Psieq}) and (\ref{Pieq}). Both constraints
cannot be met simultaneously unless $\Sigma=0$. This is only
consistent if the scalars satisfy
\begin{equation}
\sigma^\mu D_\mu \phi^i=0
\end{equation}
and
\begin{equation}
m_{3/2} = 0 \, .
\end{equation}
If $B'\ne 0$, then in order to satisfy the gravitino equation~(\ref{Pieq}), 
$\Pi$ needs to have the same phase as $\Psi$, and so
\begin{equation}
\Pi_\pm(x^\mu)= f_\pm(z,t) \Pi_\pm(r,\theta) \, .
\end{equation}
Now this is not generally consistent with eq.~(\ref{Psieq}) as 
$(\bar \sigma^t\partial_t - \bar \sigma^z \partial_z)\Pi$ does not
vanish for non-zero $\Pi$. This implies that
\begin{equation}
\Pi=0
\end{equation}
and so for consistency we also need
\begin{equation}
B'=0 \, .
\end{equation}
In conclusion, we find that the usual extension of zero mode solutions
to massless currents is only consistent in supergravity for BPS
backgrounds. In this case the only part of the gravitino which does not
vanish is $\Psi$. In the non-BPS cases, the spectrum does not
appear to contain any massless excitations, even though it does have
zero modes. A possible resolution of this is that the normalisable zero modes
must be extended to currents with a non-standard dispersion relation 
$k_t= h(k_z)$, instead of the usual $k_t=\pm k_z$. At low energy,
for small $k_z$, this suggests that $k_t \approx h_1 k_z$ where 
$\vert h_1 \vert \ne 1$. In the following subsection we find further 
indications that this might be the case.

\subsection{Two dimensional effective action}

We will now investigate the nature of the fermion currents on the
string by considering the two-dimensional effective action there. We
will just consider the low energy behaviour of fermionic
excitations around the cosmic string background (so we ignore all
fermion states apart from the zero modes and their corresponding
currents).

Starting with the BPS case (so $\Pi=\Sigma=0$) we substitute the
ansatz~(\ref{curr}) into the fermion action. We keep $f_\pm(z,t)$ as
arbitrary functions, and take $\Psi_\pm(r,\theta)$, etc. to be one of our zero
mode solutions. We obtain
\begin{equation}
-i\int dt dz (\vert \vert \psi_\mu \vert \vert^2 
+\vert \chi \vert^2 + \vert \lambda \vert^2) 
\bar f \bar \sigma^i \partial_i f
\end{equation}
where $i=z,t$ and
\begin{equation}
\vert\vert \psi_\mu\vert \vert^2= \frac{1}{2}\int dr d\theta C 
\vert \Psi \vert ^2 \, .
\end{equation}
Hence we have obtained an effective two-dimension action for a
massless fermion, $f$. Each BPS normalisable zero mode leads to a massless
current on the string world sheet. The norm used for the gravitino
is positive definite and coincides with the one used in the index
theorem~(\ref{gnorm}).

Let us repeat the same analysis in the non-BPS D-strings. We have seen
that the four-dimensional equations of motion do not lead to massless currents.
Here we investigate the breakdown of the masslessness condition at
the level of the effective action by extending the ansatz~(\ref{curr})
to include
\begin{equation}
\Sigma_\pm(x^\mu)= f_\pm(z,t) \Sigma_\pm (r,\theta) \, , \qquad
\Pi_\pm(x^\mu)= f_\pm(z,t) \Pi_\pm (r,\theta) \, .
\end{equation}
Substituting this into the action gives
\begin{equation}
-i\int dt dz \bar f N^i \partial_i f
\end{equation}
where
\begin{equation}
N^t = \bar \sigma^t \int dr d\theta e^{2B} C \left[ |\chi|^2 + |\lambda|^2 
+ \frac{1}{2}\left(|\Psi|^2 + |\Sigma+\Pi|^2 -|\Pi|^2 \right) \right]
\end{equation}
and
\begin{equation}
N^z = \bar \sigma^z \int dr d\theta e^{2B} C \left[ |\chi|^2 + |\lambda|^2 
+ \frac{1}{2}\left(|\Psi|^2 + |\Sigma-\Pi|^2 -|\Pi|^2\right) \right] \, .
\end{equation}
As $N^t\ne N^z$, we find that the action does not describe a two-dimensional
massless fermion. More precisely we find that the dispersion
relation for the fermion $f_\pm$ becomes
\begin{equation}
k_t = \pm \frac{ N^z}{ N^t} k_z \, .
\end{equation}
It would be very interesting to investigate the non-BPS zero modes
and their associated currents further. Indeed, the non-BPS case
seems to be plagued with two types of inconsistencies. First the
zero modes are not guaranteed to have a positive norm, although
residual gauge symmetries might be enough to gauge away the would-be
negative norm states. Even if the zero modes are of positive norm,
their physical relevance is not at all clear. Indeed they do not
seem to be naturally extendible to massless currents, and seem to
have two-dimensional Lorentz breaking dispersion relations. It 
could be that the normalisable zero modes in the non-BPS case are all gauge
artifacts. This would guarantee the absence of two-dimensional Lorentz
invariance breaking currents. However this seems unlikely, as a BPS
D-string (which has problem-free massless states) can be obtained as a
limiting case of a non-BPS string. More work is needed in this
direction in order unravel the physics of zero modes in the non-BPS case.

\section{Conclusion}
\label{sec:conc}

We have derived a general expression for the number of fermion zero modes
bound to a cosmic string. This index theorem is valid for a general
model, even if it includes massless fermions and gravitinos (both
possibilities being important for supergravity). The presence of
massless eigenstates in the mass matrix reduces the number of zero
modes. Physically this can be interpreted as the bound states mixing
with free massless states. The presence of gravitinos also reduces
the number of zero modes, firstly because the gravitinos are massless away from
the string and secondly because the string confines them less
effectively due to their different kinetic terms.
Since strong constraints on conducting strings arise from the
presence of vortons, reduced conductivity of supergravity strings
implies that the vorton constraints will be relaxed.

In particular for D-term strings in supergravity, we find there is one less
zero mode state than the corresponding model without gravitinos. This
implies that there is no zero mode on a cosmic string with winding
$n=1$. This is consistent with the results of ref.~\cite{rachel}.
Consequently these D-strings evade the stringent constraints on
chiral vortons which their global analogues were subject to. Higher 
winding number strings still have zero modes. It is
interesting to ask what happens to states on a $n>1$ string if it
splits into several $n=1$ strings. Does the wavefunction spread out
over the different strings, or must it decay?

There are rather different physical reasons behind the reduction in
zero modes in the D-term and F-term cases. The former appears to be a
curved space analogue of the superHiggs effect. The F-term case is
similar to the global case in the presence of supersymmetry breaking
terms. There it was found that supersymmetry breaking resulted in
the mixing of zero modes, which aided their destruction. In the case
of supergravity with F-terms, the gravitino mixes the left and right
moving fermions in the Lagrangian, resulting in the absence of zero 
modes. 

Usually fermion zero modes are confined to the string by Yukawa
couplings to the string Higgs fields. However in some cases the string
gauge fields and gravity is enough to confine the fermions (this is
only relevant for models with fermions which are massless outside
the string). The wavefunctions of such zero modes have power law decay
outside the string (as opposed to the usual exponential decay). Because of
this the corresponding currents are less effective at stabilising
string loops, and so the resulting vorton constraints are weaker than usual.

Finally one would like to include a gravitino mass for D-term string
models and take into account the effects of supersymmetry breaking.
As this is only possible for quantised Fayet-Iliopoulos terms
$\xi=p/n$ (with $p$ integer), and since the deficit angle at infinity implies
that $0<C_1= 1-p<1$, it seems that D-term strings are difficult to
construct when a gravitino mass is present.

\acknowledgments 
We are grateful to Renata Kallosh for useful discussions. ACD and CvdB
were supported by PPARC. SCD thanks Rachel Jeannerot and Marieke
Postma for useful comments, and was supported by the Swiss
Science Foundation and the Netherlands Organisation for Scientific
Research (NWO).

\end{document}